\documentclass[twocolumn,amsmath,amssymb,aps]{revtex4}
\usepackage{txfonts}
\usepackage{amssymb}
\usepackage{setspace}
\usepackage{graphicx}

\begin{document}

\title{Oblique Hanle Effect in Semiconductor Spin Transport Devices}

\author{Jing Li}
\altaffiliation{jli@udel.edu}
\affiliation{ Electrical and Computer Engineering Department,
University of Delaware, Newark, Delaware, 19716}
\author{Biqin Huang}
\affiliation{ Electrical and Computer Engineering Department,
University of Delaware, Newark, Delaware, 19716}
\author{Ian Appelbaum}
\affiliation{ Electrical and Computer Engineering Department,
University of Delaware, Newark, Delaware, 19716}

\begin{abstract}
Spin precession and dephasing (``Hanle effect'') provides an unambiguous means to establish the presence of spin transport in semiconductors. We compare theoretical modeling with experimental data from drift-dominated silicon spin-transport devices, illustrating the non-trivial consequences of employing oblique magnetic fields (due to misalignment or intentional, fixed in-plane field components) to measure the effects of spin precession. Model results are also calculated for Hanle measurements under conditions of diffusion-dominated transport, revealing an expected Hanle peak-widening effect induced by the presence of fixed in-plane magnetic bias fields. 
\end{abstract}

\maketitle
\newpage

Spin transport in semiconductors has recently been the subject of vigorous research because it opens possibilities for creating devices and circuits making use of the spin degree of freedom in addition to manipulation of the electron charge.\cite{ZUTICRMP, AWSCHALOM, FABIANAPL} There has been much presentation in the literature of ``spin-valve'' measurements, where the relative orientation between ferromagnetic ``injector'' and ``detector'' magnetization axes is controlled by an external magnetic field in the device plane, but in the past few years it has been firmly established that the only convincing, unambiguous proof of genuine spin transport is clear evidence of spin precession and dephasing (``Hanle effect'') in an out-of-plane magnetic field\cite{MONZON, JOHNSON1985, JOHNSON1988, LOU, APPELBAUM, ZUTICNATURE}. 

While a magnetic field entirely perpendicular to the device plane is the easiest geometry to analyze, it is often the case that there are in-plane magnetic field components as well.\cite{MOTSNYI} In addition, it is often desirable to control the relative injector/detector magnetization orientation with an in-plane magnetic bias field in conjunction with a purely perpendicular field.\cite{JONKERLATERAL} 

Here we show how the standard spin precession model (based on drift-diffusion theory)\cite{LOU, SPINFETTHRY} can be modified to incorporate these oblique fields, and compare the resulting calculations to experimental data from silicon spin transport devices.\cite{APPELBAUM} Because of silicon's intrinsic advantages for spintronics (small spin-orbit scattering and hyperfine interaction, and presence of a spin-degenerate conduction band leading to an exceptionally long spin lifetime\cite{ZUTICPRL, APPELBAUM, HUANG, JONKERNATPHYS}) and the difficulties in applying spin injection/detection methods developed for other semiconductors to it (caused by an indirect bandgap and propensity for interfacial alloying), confirmation of spin transport in this semiconductor using the Hanle effect is particularly important. We also use our model to analyze the expected effect of fixed in-plane magnetic fields on Hanle measurement in the widely-used diffusion-driven lateral spin transport technique measured with ferromagnetic nonlocal voltage probes.\cite{BHATTACHARYA, FUHRER, VANWEESGRAPHENE, JONKERLATERAL}

We wish to model the device spin detector output, which in linear response is proportional to the projection of final spin direction (after transport) on the measurement axis determined by detector magnetization. Under the influence of an oblique magnetic field $\vec{B}=B_z\hat{z}+B_y\hat{y}$, where $\hat{z}$ is in the direction normal to the device plane, and $\hat{y}$ is in-plane and along the injector/detector magnetization direction, spin is induced to precess around the magnetic field at frequency $\omega=g\mu_B\sqrt{B_z^2+B_y^2}/\hbar$. In cartesian $(x',y',z')$ coordinates where the magnetic field is along the $z'$ direction, the initial spin direction at the injector is $\vec{s_i}=(0, \sin\theta, \cos\theta)$, where $\theta$ is the angle between the injected spin direction in the device plane (along $\hat{y}$) and the effective magnetic field $\vec{B}$, as shown in the inset of Fig. 1 (a). After precession over transit time $t$, the final spin direction is $\vec{s_f}=(\sin\theta \sin\phi, \sin\theta \cos\phi, \cos\theta)$, where the azimuthal $\phi$ is the precession angle $\omega t$. If the injector and detector are in a parallel orientation, the contribution to detected signal from a single precessing electron spin with fixed transit time $t$ is then proportional to $\vec{s_i}\cdot \vec{s_f}=\sin^2\theta \cos\phi+\cos^2\theta$. This expression can be simplified to 

\begin{equation}\label{TRIG}
\frac{B_z^2\cos\omega t+B_y^2}{B_z^2+B_y^2},
\end{equation}

\noindent using trigonometric definitions $\sin\theta=\frac{B_z}{\sqrt{B_z^2+B_y^2}}$ and $\cos\theta=\frac{B_y}{\sqrt{B_z^2+B_y^2}}$.

Because the transit time from injector to detector for each electron is affected by random walk induced by diffusion, the expected spin signal is a sum of all the projection contributions at different arrival times, weighted by the arrival time distribution. This distribution function, which describes the spatio-temporal evolution of an ensemble of spins that all originate at the injector with the same spin orientation (at the same time), is given by the Green's function of the spin drift-diffusion equation.\cite{JEDEMA, LOU, SPINFETTHRY} Using Eq. \ref{TRIG}, the spin signal measurement should therefore be proportional to

\begin{equation}\label{ObliqueHanleEq}
\int_0^\infty \frac{1}{2\sqrt{\pi Dt}}e^{-\frac{(L-vt)^2}{4Dt}}
e^{-t/\tau_{sf}}\frac{B_z^2\cos\omega t+B_y^2}{B_z^2+B_y^2}dt.
\end{equation}

\noindent where $D$ is diffusion coefficient, $L$ is transit length, $v$ is drift velocity, and $\tau_{sf}$ is spin lifetime. 

In the case of an oblique uni-axial magnetic field with constant orientation $\theta$ and magnitude $|B|$, Eq. \ref{ObliqueHanleEq} reverts to

\begin{equation}\label{AppelbaumHanleEq}
\int_0^\infty \frac{1}{2\sqrt{\pi Dt}}e^{-\frac{(L-vt)^2}{4Dt}}
e^{-t/\tau_{sf}}(\sin^2\theta \cos\omega t+\cos^2\theta)dt.
\end{equation}

\noindent The magnetic field affects the problem only through the precession frequency $\omega=g\mu_B|B|/\hbar$, and the angle $\theta$ only determines the relative strength of the precession features since integration over the second term in parenthesis yields a constant. The $\sin^2\theta$ coefficient in the first term means that spin precession measurements in single-axis and nominally perpendicular magnetic fields are robust to (small) misalignments $\delta\theta$, with the only consequence (besides injector/detector magnetization switching induced by the in-plane component of the applied field) being a reduction in signal change by a factor of $\sin^2(\pi/2\pm\delta\theta)\approx 1-\delta\theta^2$, which is quadratically close to unity. 

\begin{figure}
  \centering
  \includegraphics[scale=0.47]{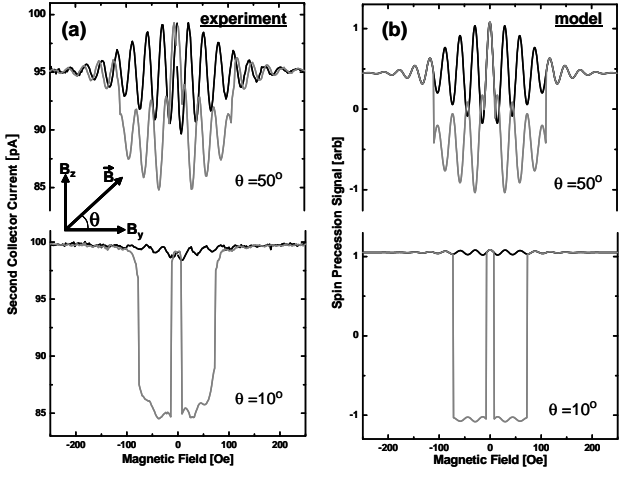}
  \caption{Comparison of experimental (a) and simulated (b) Hanle spin precession data in oblique single-axis magnetic fields using devices as described in Ref. \cite{HUANG} at 150K. Plots at the top of each panel are with magnetic field at an angle $\theta$=50$^\circ$ from the device plane and plots at the bottom of each panel are with magnetic field at an angle $\theta$=10$^\circ$. Portions of the measurement where the magnetic field magnitude increases from zero (and in-plane components switch the injector and detector magnetization) are in grey. Inset: uniaxial field geometry, where $B_y$ is in the device plane and along the injected spin direction.}
\end{figure}

\begin{figure}
  \centering
  \includegraphics[scale=0.455]{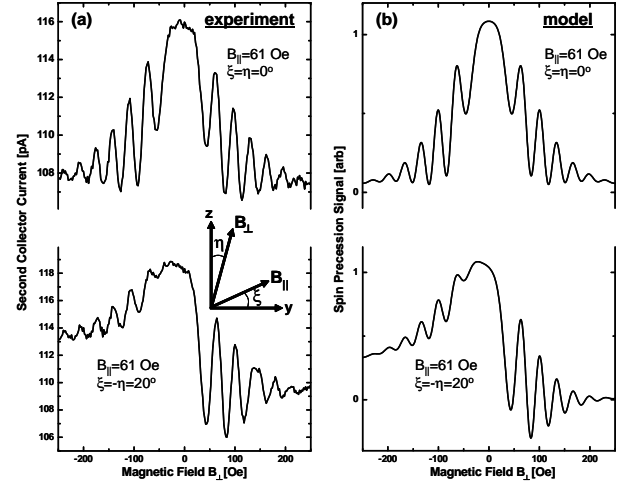}
\caption{Comparison of experimental (a) and simulated (b) Hanle spin precession measurements in a fixed magnetic field $B_{||}$ (perpendicular to the varied magnetic field $B_\perp$), for zero device misalignment (at the top of each panel) and for a 20$^\circ$ misalignment (at the bottom) in $B_{||}$=61 Oe. Inset: geometry of applied fields relative to device axes, where $z$ is the device normal and $y$ is in the device plane.}
\end{figure}

Experimental results obtained with our silicon spin-transport devices at 150K (fully described in previous work\cite{HUANG}) for $\theta=$ 50$^\circ$ and 10$^\circ$ are shown in the top and bottom of Fig. 1 (a), respectively. The spin signal in these devices is a hot-electron ``second collector current'' which has traveled ballistically through a ferromagnetic detector thin film after transport across the full thickness of an undoped single-crystal float-zone-grown Si(100) wafer. In the top and bottom of Fig. 1 (b), we have used  Eq. \ref{AppelbaumHanleEq} with $L=350\mu$m$, D=200$cm$^2$/s, $v=2.9\times 10^6$cm/s (at an accelerating voltage of 20V) and $\tau_{sf}=73$ns\cite{HUANG} to calculate the expected spin signal for corresponding magnetic field orientations. The effects of in-plane magnetization switching in the experimental data are prominent, and this is incorporated into the model results by inverting the sign at the appropriate magnetic field values. Despite the predicted (and experimentally confirmed) reduction in signal oscillation magnitude, the extrema are at identical positions regardless of the value of $\theta$. This invariance is especially important when the oscillation period is used to determine the electron spin transit time in an unintentionally misaligned magnetic field.\cite{BIQINJAP}

It is sometimes desirable to have a static in-plane bias magnetic field ($B_{||}$, in addition to a perpendicular field $B_\perp$) to control the relative injector/detector orientations.\cite{JONKERLATERAL} However, the influence of this fixed field on the measurement results cannot be ignored. In particular, because the effective field $\vec B$ forms a very small angle $\theta$ with the injected spin direction for $B_z<B_y$, the low-angle ($\phi=\omega t$) precession oscillations are suppressed. 

Experimental measurements of spin precession in a fixed in-plane magnetic field of 61 Oe are shown in the top of Fig. 2(a). As expected, the precession oscillations at small perpendicular field values are suppressed, and the shape of the signal reflects the broken up-down symmetry induced by the presence of the in-plane field.  Moreover, the oscillation extrema positions are shifted (unlike misalignment of a uniaxial field with fixed $\theta$ as shown in Fig. 1), so transit time cannot be simply deduced from oscillation period.\cite{BIQINJAP} Using Eq. \ref{ObliqueHanleEq} with $B_y=$61 Oe, we see that the model captures this behavior with high fidelity, as shown at the top of Fig. 2(b). 

Unlike in single-axis measurements, misalignments of $B_\perp$ and $B_{||}$ at angles $\eta$ and $\xi$ from $\theta=90^\circ$ ($z$ axis) and $\theta=0^\circ$ ($y$ axis), respectively, (as shown in the inset to Fig. 2 (a)) can make substantial changes to the measured Hanle spin precession signal. In this case, we have 

\begin{eqnarray}\label{MISALIGNEQ}
B_z=B_{||}\sin\xi + B_\perp \cos\eta \nonumber \\
B_y=B_{||}\cos\xi + B_\perp \sin\eta.
\end{eqnarray} 

\noindent The even symmetry with respect to the varied field ($B_\perp$) in Eq. \ref{ObliqueHanleEq} is then broken, and asymmetric Hanle curves can be obtained.

The bottom of Fig. 2(a) again shows experimental Hanle results in a fixed magnetic field $B_{||}$ of 61 Oe. However, even if $B_\perp$ and $B_{||}$ are truly orthogonal so that $\eta=-\xi$, sample misalignment (here at 20$^\circ$) has significant influence on the observed Hanle measurement, causing an obvious asymmetry in magnetic field polarity as expected. Calculations of Eq. \ref{ObliqueHanleEq} using Eqs. \ref{MISALIGNEQ} as a function of $B_\perp$ with the same transport parameters used above, and shown in the bottom of Fig. 2(b), agree quite well with this behavior. Although the exaggerated 20$^\circ$ misalignment used here is unlikely to be unintentional, even small misalignments (which are unavoidable in practice) can leave distinct signatures on the experimental data if a fixed in-plane bias magnetic field is used.

\begin{figure}
  \centering
  \includegraphics[scale=0.9]{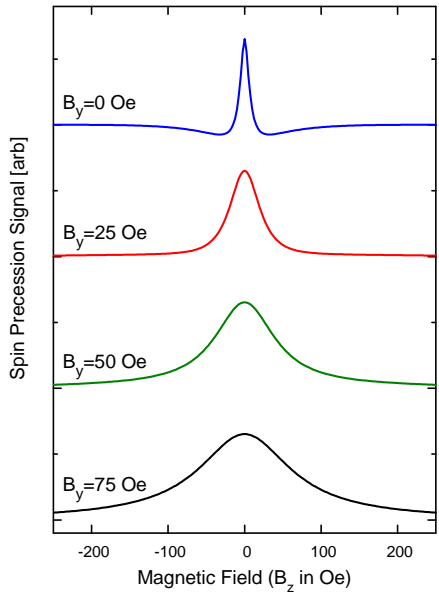}
  \caption{Simulated spin precession measurements for diffusion-driven transport across 1$\mu$m, using Eq. \ref{ObliqueHanleEq} as a function of perpendicular magnetic field $B_z$, with a constant magnetic field ($B_y$) of 0, 25, 50, and 75 Oe applied in the device plane. Simulation results are offset for clarity.}
\end{figure}

The strong influence of fixed in-plane fields seen here with drift-dominated spintronics devices\cite{DEPHASING} are also expected in the diffusion-dominated regime. Evaluating Eq. \ref{ObliqueHanleEq} as a function of $B_z$ with parameters relevant for nonlocal spin-valve devices at low temperature and high doping ($D=1$cm$^2$/s, $L=1\mu$m, $v=0$, and $\tau_{sf}=10$ns) we obtain results for $B_y$ = 0, 25, 50, and 75 Oe as shown in Fig. 3 from top to bottom. Notably, secondary oscillations at non-zero $B_z$ (most evident in experiments where drift is the dominant transport mechanism)\cite{LOU,APPELBAUM} are not seen here because of the wide arrival-time distribution driven by diffusion and the consequently strong spin dephasing.

The most salient feature of our model results in Fig. 3 is that as the in-plane magnetic bias field $B_y$ increases, the central Hanle peak increases in width. This can be heuristically understood by considering that when $B_z<B_y$, the spin direction has a positive projection on the measurement axis regardless of precession angle $\phi$. Therefore, to cause the same amount of dephasing from signal cancellation, $B_z$ must be increased as $B_y$ increases.

Incorporating the necessarily non-zero lateral width of the injector and detector introduces a constant source of systematic spin dephasing by adding a fixed transit-length uncertainty that is not expected to affect this trend. Increasing the diffusion constant $D$ and decreasing the spin lifetime $\tau_{sf}$ reduces the strength of this peak-widening effect, but the general peak-widening behavior seen here persists because of its geometric origin. However, the presence of magnetic and electric-field inhomogeneities in the transport path will modify the measured lineshape and could partially obscure the expected trend. 

In summary, the Hanle spin precession signal magnitude is reduced by a factor of $\sin^2\theta$ when making measurements in single-axis oblique magnetic fields at an angle $\theta$ to the device plane. Besides in-plane magnetization switching of injector and detector, small misalignments therefore cause no modification to precession measurements to first order. When a fixed in-plane magnetic field is used in conjunction with a  varied perpendicular field, Hanle measurements are affected by a suppression of precession oscillations at low fields and asymmetry when the device is misaligned. Results from simulations of devices where diffusion is the dominant transport mechanism indicate that a widening of Hanle peak width is expected in the presence of these in-plane bias fields.

Support from DARPA/MTO and ONR is acknowledged.

\end{document}